# Fund Characteristics and Performances of Socially Responsible Mutual Funds: Do ESG Ratings Play a Role?

**Nandita Das**
College of Business
Delaware State University, Delaware DE 19901
Email: ndas@desu.edu

**Bernadette Ruf**
College of Business
Delaware State University, Delaware DE 19901
Email: bruf@desu.edu

**Swarn Chatterjee**
University of Georgia, Athens GA 30602
Email: swarn@uga.edu

**Aman Sunder**
College for Financial Planning
Centennial CO 80112
Email: aman.sunder@cffp.edu

## ABSTRACT

This paper examines the risk-adjusted performance and differential fund flows for socially responsible mutual funds (SRMF). The results show that SRMF rated high on ESG, perform better than lower rated ESG funds during the period of economic crisis. The findings also show that low ESG rated SRMF had higher differential cash-flows than high rated ESG funds except for the period of economic down turn. The findings are of interest to financial advisors, investors, mutual fund managers, and researchers on how SRMF performance responds to periods of economic downturn and expansion.

## INTRODUCTION

Do socially responsible mutual funds (SRMF) that rate higher on social responsibility score have higher risk-adjusted returns compared to SRMF funds with a lower rating? Do higher rated SRMF receive more cash inflows than SRMF with a lower rating? Is the relationship between risk-adjusted return and social responsibility score consistent over different economic cycles? This study attempts to address these questions.

According to US SIF (2016), investment in the US-based SRMF has grown over the past two years by 33% with assets under management totaling $8.72 trillion at the beginning of 2016. Prior academic research by Nofsinger and Varma (2014), Renneboog, Horst, and Zhang, (2008), and Bollen (2007) on SRMF has focused primarily on comparing the performance of conventional mutual funds with SRMF. Renneboog et al. (2008) hypothesize that the superior performance of SRMF funds compared to conventional funds can be attributed to the fund-portfolio composition. An explanation could be that the

SRMF funds invest in companies that demonstrate corporate social responsibility and transparency of their operations. It is expected that these companies are likely to be better managed and hence, generate better risk-adjusted performance. While this conjecture has produced some mixed findings, Rathner (2013) in a meta-analysis study, find no statistical difference in performance between US-domiciled SRMF and conventional funds. However, Nofsinger and Varma (2014), find that SRMF outperformed conventional funds during periods of market crisis and underperformed during non-crisis periods.

Very little research has been conducted on the determinants of performance and fund flow among SRMF funds. Given the recent growth in the SRMF, there is a need to explore the characteristics of SRMF that explain the risk-adjusted performance within the SRMF universe. This study investigates whether SMRF social performance ratings translate into higher risk-adjusted returns. Based on the Morningstar® Portfolio ESG Score™ (ESG Score) (Justice & Hale, 2016), this study categorizes the SRMF into low, mid, and high ESG groups. Specifically, this study examines whether socially responsible funds with high ESG ratings outperform the lower ESG rated funds on a risk-adjusted basis.

The study also examines whether the differential flow of SRMF can be explained solely by the assigned ESG rating controlling for past fund performance on the risk-adjusted basis, and other important factors. The findings from this study will build on the current understanding of how the social-responsibility ratings of funds relate to fund performance and differential fund flow for the SRMF.

## LITERATURE REVIEW

### Financial Performance

**Risk-Adjusted Return.** Previous researchers find no difference in risk-adjusted performance of SRMF and conventional funds using different social responsible rating systems. The studies include Hamilton, Jo, and Statman (1993), Statman (2000), Bello (2005), Shank, Manullang, and Hill (2005), Gil-Bazo, Ruiz-Verdú, and Santos (2010), Renneboog et al. (2008) and Rodríguez (2010). Gil-Bazo et al. (2010) conclude that SRMF have a superior risk-adjusted performance compared to conventional funds before and after adjusting for costs. They also found that when SRMF were managed by companies specializing in social responsibility, SRMF outperformed conventional funds. On the contrary, Chang and Witte (2010) study find no difference in risk-adjusted performance between SMRF and conventional funds except two categories: balanced funds and fixed-income funds.

**Fund Flow.** As to the relationship between fund flow and fund performance, Bollen (2006) concludes that monthly fund flow volatility of SRMF is lower than conventional funds. Cash-flows into SRMF are more sensitive (less sensitive) to lagged positive (negative) performance than are cash-inflows into conventional funds. Furthermore, the study finds the difference between SRMF and conventional funds was robust over time and persisted as funds age. In a similar study, Renneboog et al. (2008) investigate the effect of *smart money*, i.e., whether investors select funds that generate superior performance in subsequent periods. They find no difference in the alphas for the inflow and outflow portfolios of SRMF and conventional funds domiciled in the United States. They also find that alphas for inflow portfolios are not significantly different from zero, suggesting that investors are unable to select funds that would outperform the benchmark factors in the future, but can identify funds with poor performance. Nofsinger and Varma (2014) find that SRMF assets under management (AUM) increased by more than 13 percent during 2007-2009 while conventional funds remained relatively flat. SRMF also outperformed the conventional funds during the market crisis but underperformed the conventional funds during normal economic cycles (non-crisis periods). The authors conclude that this difference can be attributed to the socially responsible characteristics of the underlying stocks within the portfolio rather than the stock picking ability or portfolio management skills of SRMF fund managers.

**Investor Motivation.** Studies from household finance have found that cultural differences among households play a role in many investors' decision to prefer safety and more stable forms of investments (Chatterjee & Zahirovic-Herbert, 2014; 2011; Kim et al., 2012). Investors' motivation to invest in SRMF has increased over the recent years. Studies by Bollen (2007) and Statman (2005) posit that investments in SRMF have become popular because investors who believe in stewardship of environmental responsibility, socially responsible behavior, and ethical corporate governance and leadership derive utility from investing in socially responsible funds. For these investors, the utility derived from making investments consistent with their ethical and moral values sometimes supersedes the utility generated purely based on a profit motive.

**Investor Characteristics**. Researchers have examined the characteristics of investors who participate in socially responsible investing. Nilsson (2008) identifies three types of investors that participate in the SRMF. The first type of investor is solely motivated by mandate of social responsibility. The second type of investor is motivated by the investment return of the funds and only interested in SRMF if they can earn a higher return than conventional funds. The third type of investor has a dual objective—a social responsibility mandate and a profit motive.

## METHODOLOGY

### Conceptual Framework

The conceptual framework for this study is developed based on prior SRMF studies. Nofsinger and Varma (2014) find that SRMF funds outperform conventional funds during periods of higher market volatility but probably underperform the market during periods of non-crisis. It is therefore expected that the high ESG rated mutual funds will similarly outperform the low ESG rated mutual funds during periods of economic crisis but underperform the low ESG rated mutual funds under normal market conditions. This dampening effect on the downside risk for high ESG rated funds is likely due to the type of companies the funds held within their portfolios. The underlying stocks of the companies in the SRMF with high ESG score are less likely to have exposure to extreme negative events because of the environment, social, and governance mandates in these companies. For example, the underlying portfolios of high ESG score SRMF funds will have lower exposure to costs arising from irresponsible stewardship of the environment, poor stewardship of financial resources, or from stakeholder lawsuits. According to Mcguire, Sundgren, and Schneeweis (1988), the socially responsible firms held within the portfolios of high ESG rated funds will likely have more positive associations with the respective communities and regulators in the industries within which they operate. Additionally, Verwijmeren and Derwall (2010) find that firms that provide greater employee satisfaction are less likely to exhibit bankruptcy risks.–As a result, the performances of socially responsible stocks (and the mutual funds holding these securities) are expected to be stable and less risky during periods of regular market activity as well as during periods of market crisis.

The Psychological Attraction Theory of financial regulations by Hirshleifer (2007) posits that the government and the policymakers are more likely to impose stronger regulations in the marketplace reactively rather than proactively in response to a crisis. Shefrin and Statman (1993) argue that although the qualities of lower risk and higher standards of responsibility are practiced by the socially responsible companies (and by default in funds holding these securities) during all economic cycles, the advantages of this low-risk and socially responsible mandate are noticed by stakeholders (investors, government, and policy makers) during periods of market uncertainty. It is therefore expected that differential flows into highly rated socially responsible funds would also increase during periods of market uncertainty. The government, policymakers, and other important stakeholders are more likely to come up with regulations negatively affecting the bad actors immediately following an economic crisis. The socially responsible firms that self-impose a higher standard of responsibility, are less likely to be negatively affected by any new regulation that might be brought into the industry. Hence, socially responsible corporations are expected to be penalized less during periods of the market downturn than other firms.

According to the Prospect Theory by Kahneman and Tversky (1979), investors' perceived decline in utility from a loss is higher than their increase in perceived utility for an equivalent amount of gain. Therefore, the risk averse investors will be willing to sacrifice some return during normal economic times and favor of more stable performances in their portfolios during periods of market uncertainty. Furthermore, Cox, Brammer, and Millington (2004) find that institutional portfolios managers prefer investing in corporations with strong corporate governance because of their stable performance related characteristics.

**Hypotheses**

Based on the conceptual framework and the findings from previous literature, this study hypothesizes that higher ESG rated funds have higher risk-adjusted return compared to lower ESG rated funds. This study also expects higher differential flows into the higher ESG rated funds compared to lower ESG rated funds because of the greater expected stability of these funds over all periods and because these funds will be more attractive to investors with a socially responsible mandate.

$H_1$: The lower ESG rated SRMF will underperform the higher ESG rated funds after controlling for other fund related characteristics.

$H_2$: The lower ESG rated SRMF will receive lower differential flow than higher ESG rated SRMF after controlling for other fund related characteristics.

**Data Selection**

The data is from the Morningstar® database and covers the time-period from 2005 to 2016. The funds are selected based on the following criteria: had a socially responsible mandate; domiciled in the United States; and surviving funds from 2005 to 2016 that were rated in the top half of Morningstar® Sustainability Rating™. The Morningstar® Portfolio ESG Score™ *(ESG Score)* is used to categorize the SRMF in this study. Based on these screening criteria the total sample narrowed down to 73 SRMF that survived the entire study period (2005-2016). Thus, the study consists of 12 years of monthly data. The study divides the study-period into 3 sub-periods 2005-2008 (the period before and during the Great recession), 2009-2012 (the period of recovery immediately after the Great Recession), and 2013-2016 (period of economic expansion).

Morningstar® database includes the ESG ratings for funds that incorporates environmental, social, and governance factors. Environmental issues include climate change and carbon emissions, air and water pollution, energy efficiency, water scarcity, waste management and deforestation. Social issues include product safety, data protection/privacy, gender and diversity, employee engagement, supply chain management, and labor standards. Governance issues include board composition, audit committee structure, executive compensation, lobbying, political contributions and bribery and corruption.

This study divides the funds into tertile, where funds in top 33% of the assigned ESG scores are categorized as High ESG, followed by funds in the middle 33% of ESG scores which are categorized as Mid ESG, and funds in the lowest 33% of ESG scores are categorized as Low ESG. Binary variables are created to represent each tertile as a variable in the empirical model.

**Empirical Analyses**

The study uses ordinary least squares (OLS) regression models to understand the impact of SRMF fund ratings on the risk-adjusted performance and differential flow of investments into SRMF funds after controlling for fundamental factors. The fundamental factors used in the study are management tenure, expense ratio, age, and fund size. The study carries out this analysis for the complete study period (2005-2016) and the three sub-periods (2005-2008, 2009-2012, and 2013-2016). The control variables included in the empirical models have been found to be associated with fund performance in previous literature. To examine whether ESG ratings are significantly associated with monthly returns among the SRMF funds, the study regresses the rolling average of the Sharpe ratio against fundamental factors. The regression model is as follows:

$$SHARPE_i = \alpha_i + \beta_{mid_esg} Mid_ESG_i + \beta_{low_esg} Low_ESG_i + \beta_{tenure} TENURE_i + \beta_{exp} EXP_i + \beta_{size} SIZE_i + \beta_{age} AGE_i + \varepsilon_i \quad (1)$$

Where:

$SHARPE_i$ is the rolling average Sharpe ratio of the mutual funds. Sharpe ratio is calculated as follows:

$$SHARPE_i = \frac{(\mathrm{R_p} - \mathrm{R_f})}{\sigma_\mathrm{p}} \quad (2)$$

where:

$R_p$ is the return of the mutual fund

$R_f$ is the risk free rate

$\sigma_\mathrm{p}$ is the standard deviation of the portfolio

$Mid_ESG_i$is a binary variable for funds in the middle 33% of the ESG ratings, and $Low_ESG_i$ are the funds in the lowest 33% of the ESG ratings (compared against the reference group of funds in the highest 33% of the ESG ratings, $High_ESG_i$).

$TENURE_i$represents the control for management tenure. Previous studies by Kostovetsky (2017), Amihud and Goyenko (2013), Lin and Yung (2004), find management tenure to be a predictor of mutual fund performance.

$EXP_i$ is the expense ratio of the funds. Tufano and Sevick (1997), Del Guercio, Dann, and Partch (2003), Malhotra, Jaramillo, and Martin (2011) find that better managed funds have lower expense ratios.

$SIZE_i$ the fund size, is included as a control variable because previous studies by Berk and Green (2004), and Bauer, Koedijk, and Otten (2005) find that fund size is associated with both fund performance and cash-flows in mutual funds.

$AGE_i$ is the control for fund's age. Chevalier and Ellison (1999) and Grinblatt and Titman (1989) find age to be associated with mutual fund performance.

Investor sentiment is measured as the level of differential flow. Under status quo, every SRMF should experience an inflow (outflow) of funds that is proportional to the percentage of the SRMF industry assets that the fund owns. This is the theoretical flow for the SRMF fund. In short, with nothing changing, the study expects the SRMF fund's representation in the industry, in terms of it share to remain fixed. Differential flow is the extra flow of funds that an SRMF fund receives over and above its theoretical flow (proportional flow). The differential flow is calculated as a difference of real flow and theoretical flow divided by theoretical flow. The model is expressed using the following equations:

$$\mathrm{FL}_\mathrm{t}^\mathrm{i} = \mathrm{AUM}_\mathrm{t}^\mathrm{i} - \mathrm{AUM}_\mathrm{t-1}^\mathrm{i}\left(1 - \mathrm{r}_\mathrm{t}^\mathrm{i}\right) \quad (3)$$

$$\mathrm{FL}_\mathrm{t}^\mathrm{i} = \mathrm{AUM}_\mathrm{t}^\mathrm{i} - \mathrm{AUM}_\mathrm{t-1}^\mathrm{i}\left(1 - \mathrm{r}_\mathrm{t}^\mathrm{i}\right) \quad (4)$$

$$\overline{\overline{\mathrm{FL}}}_\mathrm{t}^\mathrm{i} = \frac{\mathrm{AUM}_\mathrm{t-1}^\mathrm{i}}{\mathrm{AUM}_\mathrm{t-1}^\mathrm{Agg}} \times \mathrm{FL}_\mathrm{t}^\mathrm{Agg} \quad (5)$$

$$\mathrm{AUM}_\mathrm{t-1}^\mathrm{Agg} + \sum_\mathrm{i=1}^\mathrm{N}(\mathrm{AUM}_\mathrm{t-1}) \quad (6)$$

$$FL_i^{Agg} \quad \sum_{i=1}^{N}(FL_t)_i \tag{7}$$

$$Diff(FL_t^i) = \frac{FL-\overline{\overline{FL}}_t^i}{FL_t^i} \tag{8}$$

where

$FL_t^i$ is the real flow for SRMF fund *i* in time-period *t,*
$AUM_t^i$ represents assets under management for SRMF category *i* in time-period *t,*
$AUM_{t-1}^i$ is the assets under management for SRMF category *i* in previous time-period *t-1*,
$r_t^i$ is the return SRMF Category *i* in time-period *t,*
$\overline{\overline{FL}}^i$ is the theoretical flow for SRMF Category *i* in time-period *t,*
$AUM_{t-1}^{Agg}$ is the assets under management for all the SRMF Categories in the database (a good proxy for the SRMF industry size) for time-period *t-1.*
$FL^{Agg}$ is flow of funds (new money) to the SRMF industry for time-period *t,* and $Diff(FL_t^i)$ is the differential flow of funds to SRMF Category *i* in time-period *t,*
Equations 3 through 7 describe the model that the study uses to calculate investor sentiment. When $Diff(FL_t^i)$is positive there is a net inflow of excess funds and when it is negative the SRMF Category experienced an outflow of funds. It is important to realize that this study is not interested in measuring inflow or outflow. The study measures excess inflow or outflow experienced by that category.

The regression model is as follows:

$$Diff(FL^i{}_t) = \alpha_i + \beta_{return}[R_{i,t-1}] + \beta_{mid_esg}[D_{mid_esg_{i,}}] + \beta_{low_esg}[D_{low_esg_{i,}}] + \beta_{tenure}[TENURE_i] + \beta_{exp}[EXP_{i,}] + \beta_{size}[SIZE_i] + \beta_{age}[AGE_i] + \varepsilon_{pt} \tag{9}$$

where:

$Diff(FL_t^i)$ is the differential flow of funds to SRMF fund *i* in year *t,*
$R_{i,t-1}$Ri,t is the monthly return of the fund lagged by 1 period;
$D_{mid_esg_{i,}}$is a binary variable for funds in the middle 33% of the ESG ratings and $D_{low_esg_{i,}}$are the funds in the lowest 33% of the ESG ratings (compared against the reference group of funds in the highest 33% of the ESG ratings);
$TENURE_i$ is the management tenure;
$EXP_{i,}$is the expense ratio of the funds;
$SIZE_i$ is the fund size; and
$AGE_i$is the fund's age
The above empirical models (Sharpe Ratio and Differential Flow) from equations 1 through 8 are run for the 2005-2016; 2005-2008; 2009-2012; and 2012-2016 periods.

## RESULTS

### Descriptive Statistics

Descriptive statistics are provided in Table 1 and broken down into Panels reflecting the complete study period of 144 months and the three sub-periods of 48 months each.

**Time-period 2005-2016 (Panel A).** During the 2005-2016 period, the monthly returns are significantly higher for Low ESG funds (0.65; $p<0.01$) than for Mid ESG and High ESG funds. However, the risk-adjusted returns, as shown by the Sharpe ratio, are significantly higher for Mid ESG funds than for High ESG funds (0.76; $p<0.001$). The Mid ESG funds also have significantly lower expense ratios compared to the Low ESG and High ESG funds (1.09; p<0.001). However, Low ESG funds received the most monthly cash inflow during this period than the other ESG fund categories.

**Time-period 2005-2008 (Panel B).** The average return of funds during the 2005-2008 period, which included the 2008 market crash, is negative for all ESG fund categories. The High ESG funds have the lowest average negative monthly return (-0.32%; p<0.001) and the highest Sharpe Ratio during this period (0.56; p<0.001) compared to the Low ESG and Mid ESG funds. The Mid ESG funds have the lowest expense ratios (1.09; p<0.001), and the Low ESG funds have the highest fund flow during this period ($129,944; p<0.001) compared to the other ESG fund categories.

**Time-period 2009-2012 (Panel C).** The 2009-2012 period witnessed a slow economic recovery from the stock market crash of 2008. The Low ESG funds have the highest average monthly return compared to the other fund categories during this period (1.33%; p<0.001). The Low ESG funds also have the highest Sharpe ratio, although the chi-square test results for Sharpe ratios are not significant. The Mid ESG funds have the lowest expense ratios (1.12; p<0.001) and received the highest monthly cash-flow ($819,636; p<0.001) compared to Mid ESG and High ESG funds during this period.

**Time-period 2013-2016 (Panel D).** The US economy witnessed strong economic recovery during the 2013-2016 period. Similar to the prior study period, the Low ESG funds have the highest average return during this period (0.99; p<0.001). However, Mid ESG funds have the highest average Sharpe ratio (1.26; p<0.001) and the lowest expense ratios (1.08; p<0.001) compared to the Low ESG and High ESG funds during this period. Consistent with the prior time-period, the Low ESG funds received the most cash inflow ($688,974; p<0.001) compare to Mid ESG and High ESG funds.

In summary, across the three time-periods, monthly returns were significantly higher for Low ESG funds than the other funds, except during the great recession. Depending on the time-period examined, the results on the Sharpe ratio varied during the great recession, High ESG funds performed significantly better than the other fund categories. During the beginning of the recovery, there were no significant differences in performance among the fund categories. In the later period when recovery was stronger, Mid ESG funds performed significantly better than the other fund categories. For all three time-periods, Mid ESG had the lowest expense ratio. Low ESG funds received the most positive differential cash-flows for all three periods except for 2005-2008 when it had a higher positive cash-flow of $1,297,944 compared to High ESG funds with $1,270,021.

## Risk-Adjusted Returns

Table 2 presents the results of regression run on the Sharpe Ratio as the dependent variable. The Sharpe Ratio is a measure of risk-adjusted return. The regression results were divided into four different panels: Panel A represented the time-period of the entire study between the years 2005 and 2016; Panel B represented the *pre-recession and through the great recession* time-period between the years 2005 and 2008; Panel C represented the *post-recession* time-period between the years 2009 and 2012; and Panel D represented the time-period between the years 2013 and 2016, the time of economic expansion.

**Time-period 2005-2016 (Panel A).** During the overall 2005-2016 period, Low ESG funds have significantly higher risk-adjusted return when compared with the reference group of SRMF in High ESG. Similarly, management tenure is positively associated with risk-adjusted returns of the funds. The expense ratios and fund size are significant and negatively associated with the risk-adjusted return of the funds.

**Time-period 2005-2008 (Panel B).** During the 2005-2008 period, Low ESG funds and Mid ESG funds have significantly lower risk-adjusted return when compared with the reference group of High ESG. Management tenure is positively associated with the risk-adjusted return of the funds during this period. The expense ratio is significant and negatively associated with risk-adjusted returns of the funds during this period.

**Time-period 2009-2012 (Panel C).** During the 2009-2012 period, Low ESG and Mid ESG funds have significantly higher risk-adjusted returns when compared High ESG funds. The expense ratios and fund size were significant and negatively associated with risk-adjusted returns of the funds during this period.

**Time-period 2013-2016 (Panel D).** During the 2013-2016 period, Low ESG and Mid ESG funds had significantly higher risk-adjusted returns when compared with the reference group High ESG. The expense ratios and fund size are significant and negatively associated with risk-adjusted returns of the funds during this period.

## Fund Flows

Table 3 presents the results of regression run on the differential flow as the dependent variable.

**Time-period 2005-2016 (Panel A).** The results show that during the overall 2005-2016 period, Low ESG funds have significantly higher fund flows and Mid ESG have significantly lower fund flows when compared with the reference group of High ESG funds. The monthly return in the previous period is significant and positively associated with fund flows in the subsequent period. Similarly, management tenure is also positively associated with positive fund flows. The expense ratio is significant and negatively associated with fund flows during this period.

**Time-period 2005-2008 (Panel B).** During the 2005-2008 period, the fund flows for the Low ESG and Mid ESG funds did not significantly differ from the High ESG fund flows. The monthly return in the previous period is significant and positively associated with fund flows in the subsequent period. Similarly, fund size is positively associated with positive fund flows.

**Time-period 2009-2012 (Panel C).** The results indicate that during the 2009-2012 period, the Low ESG funds have significantly higher fund flows and the Mid ESG funds have significantly lower fund flows when compared with the reference group of High ESG funds. The monthly return in the previous period is significant and positively associated with fund flows in the subsequent period. Similarly, management tenure is positively associated with positive fund flows. The expense ratio is significant and negatively associated with fund flows during this period.

**Time-period 2013-2016 (Panel D).** During the 2013-2016 period, the results show that Low ESG funds have significantly higher fund flows and Mid ESG funds have significantly lower fund flows when compared with the reference group of SRMFs High ESG funds. The monthly return in the previous period is significant and positively associated with fund flows in the subsequent period. Similarly, management tenure is positively associated with positive fund flows. The expense ratio is significant and negatively associated with fund flows during this period.

## DISCUSSION and CONCLUSION

Using the 2005-2016 SRMF data from Morningstar® the study examines the risk-adjusted performances and differential flows across the highest rated, medium rated, and lower rated ESG funds. The results indicate that compared to the highest rated ESG funds, the Sharpe ratios for medium and lowest rated mutual funds are higher over the 2009-2016 period. However, in the 2005-2008 panel that included the period of severe market downturn, the lowest and the medium rated SRMF funds have a lower Sharpe ratio than the highest rated SRMF funds. This finding is consistent with the hypothesis that lower rated SRMF funds under performed the highest rate SRMF during periods of market downturn. These findings are similar to the findings of Nofsinger and Varma (2014) that the SRMF outperformed the conventional mutual funds during periods of market uncertainty but underperformed during other times of normal economic activity.

The findings partially support the second hypothesis that differential flow will be lower for the lower ESG rated SRMF when compared with the higher rated SRMF because investors preferred the stability and lower volatility that socially responsible funds offer. The results show that medium rated ESG funds had a lower differential flow, but the SRMF funds in the lowest tertile had higher differential flows compared to the SRMF in the highest tertile of ESG scores. The study finds that when compared with the SRMF in the top tertile of ESG scores, the lower rated funds have higher differential flow and medium rated SRMF funds have lower differential fund flows across all the periods except for the 2005-2008 period. More research is needed to understand the reason for this difference across the periods. This finding corroborates with the findings from Nilsson (2008) that the SRMF investments attract different types of investors. Investors who value socially responsible mandates over profit motive were more likely to continue their investments into highly rated ESG funds, while investors who were driven mainly by a profit motive were more likely to invest money into the lowest rated ESG funds, which had the highest past returns during the periods of the study. The exception to this trend was observed during the 2005-2008 period when the markets fell significantly at the end of 2008. It is possible that many investors simply stayed away from investing during this period. Overall, the results corroborate with the findings from Bollen (2007) that differential flows are associated with the fund's risk-adjusted performance in the previous period.

Furthermore, the study finds evidence that mutual fund expense ratios are negatively associated with risk-adjusted returns across the entire period of the study. These findings are consistent with Del Guercio et al. (2003) and Malhotra et al. (2011) that expense ratio is negatively associated with fund performance. Further, the study indicates that expense ratio is also negatively associated with fund flows in the post-recession period. It is possible that due to the downturn in financial markets, and greater media attention on performances of securities in the periods immediately following recession many investors were sensitive to costs and therefore avoided investing fresh money into the more expensive mutual funds with higher expense ratios. Among other variables, the study finds that fund size is negatively associated with differential flow during most of the study period, but the association between fund size and differential flows is positive during the period of great recession, thus indicating that during periods of market uncertainty, investors possibly look to fund size as a signal of quality. Further research is needed in the future to investigate this association. The finding that the highest ESG rated funds hold up better during conditions of volatility should be of interest to the financial advisors and financial planners, as a potential asset class for diversification, during periods of market uncertainty.

## Tables

Table 1: Descriptive Statistics

**Year 2005-2016 (Panel A)**

| ESG Category | Monthly Return | Sharpe Ratio | Expense Ratio | Fund Flow |
|---|---|---|---|---|
| Low ESG | 0.65 | 0.71 | 1.46 | $934,520 |
| Mid ESG | 0.56 | 0.76 | 1.09 | $(15,000,000) |
| High ESG | 0.51 | 0.65 | 1.68 | $(470,440) |
| Overall | 0.58 | 0.70 | 1.42 | $(4,584,270) |
| Chi-Squared | **p<0.01 | ***p<0.001 | ***p<0.001 | ***p<0.001 |

**Year 2005-2008 (Panel B)**

| ESG Category | Monthly Return | Sharpe Ratio | Expense Ratio | Fund Flow |
|---|---|---|---|---|
| Low-ESG | -0.39 | 0.30 | 2.02 | $1,297,944 |
| Mid-ESG | -0.37 | 0.40 | 1.08 | $ (17,000,000) |
| High-ESG | -0.32 | 0.56 | 1.66 | $1,270,021 |
| Overall | -0.36 | 0.42 | 1.60 | $ (4,487,747) |
| Chi-Squared | ***p<0.001 | ***p<0.001 | ***p<0.001 | ***p<0.001 |

**Year 2009-2013 (Panel C)**

| ESG Category | Monthly Return | Sharpe Ratio | Expense Ratio | Fund Flow |
|---|---|---|---|---|
| Low-ESG | 1.33 | 0.63 | 1.23 | $ 819,636 |
| Mid-ESG | 1.12 | 0.60 | 1.12 | $ (16,900,000) |
| High-ESG | 1.11 | 0.43 | 1.71 | $ (1,483,530) |
| Overall | 1.19 | 0.56 | 1.36 | $ (5,546,462) |
| Chi-Squared | ***p<0.001 | p<0.347 | ***p<0.001 | ***p<0.001 |

**Year 2013-2016 (Panel D)**

| ESG Category | Monthly Return | Sharpe Ratio | Expense Ratio | Fund Flow |
|---|---|---|---|---|
| Low-ESG | 0.99 | 1.18 | 1.11 | $688,794 |
| Med-ESG | 0.94 | 1.26 | 1.08 | $(11,300,000) |
| High-ESG | 0.75 | 0.96 | 1.66 | $ (1,176,057) |
| Overall | 0.89 | 1.13 | 1.29 | $ (3,716,703) |
| Chi-Squared | ***p<0.001 | ***p<0.001 | ***p<0.001 | ***p<0.001 |

Table 2: Regression Results - Risk Adjusted Performance

| **Sharpe Ratio** | **Panel A**<br>Yr. 05-16<br>Coeff.<br>(SE) | **Panel B**<br>Yr. 05-08<br>Coeff.<br>(SE) | **Panel C**<br>Yr. 09-12<br>Coeff.<br>(SE) | **Panel D**<br>Yr. 13-16<br>Coeff.<br>(SE) |
|---|---|---|---|---|
| ESG Scores (*Ref: High ESG*) | | | | |
| Mid ESG | 0.01<br>(0.85) | -0.04*<br>(0.02) | 0.04**<br>(0.01) | 0.03*<br>(0.02) |
| Low ESG | 0.04***<br>(0.00) | -0.04*<br>(0.02) | 0.06***<br>(0.00) | 0.05*<br>(0.03) |
| Management Tenure | 0.004*<br>(0.002) | 0.004***<br>(0.00) | 0.00<br>(0.78) | 0.00<br>(0.68) |
| Expense Ratio | -0.01*<br>(0.008) | -0.03*<br>(0.01) | -0.03**<br>(0.01) | -0.03*<br>(0.01) |
| Fund Size | 0.02***<br>(0.00) | 0.16<br>(0.71) | 0.00<br>(0.65) | 0.02***<br>(0.00) |
| Age | 0.01<br>(0.10) | 0.04<br>(0.92) | 0.08<br>(0.42) | 0.03<br>(0.31) |
| Intercept | 0.52***<br>(0.00) | 1.10***<br>(0.00) | 1.19***<br>(0.00) | 0.89***<br>(0.00) |

****p<0.001; **p<0.01; *p<0.05*

Table 3: Regression Results - Differential Flow

| Differential Flow | Panel A<br>Yr. 05-16<br>Coeff.<br>(SE) | Panel B<br>Yr. 05-08<br>Coeff.<br>(SE) | Panel C<br>Yr. 09-12<br>Coeff.<br>(SE) | Panel D<br>Yr. 13-16<br>Coeff.<br>(SE) |
|---|---|---|---|---|
| Lag_1 Monthly Return | 0.34***<br>(0.01) | 0.16***<br>(0.01) | 0.52***<br>(0.03) | 0.57***<br>(0.03) |
| Mid ESG | -0.02**<br>(0.01) | 0.00<br>(0.01) | -0.05**<br>(0.02) | -0.03***<br>(0.01) |
| Low ESG | 0.04***<br>(0.01) | 0.01<br>(0.01) | 0.06***<br>(0.02) | 0.04***<br>(0.01) |
| Tenure | 0.00***<br>(0.00) | 0.00<br>(0.00) | 0.00**<br>(0.00) | 0.00***<br>(0.00) |
| Expense Ratio | -0.02<br>(0.01) | 0.00<br>(0.00) | -0.05***<br>(0.01) | -0.03***<br>(0.01) |
| Fund Size | -0.02***<br>(0.00) | 0.00***<br>(0.00) | -0.02***<br>(0.09) | -0.03***<br>(0.00) |
| Age | 0.01<br>(0.01) | 0.00<br>(0.00) | 0.00<br>(0.10) | -0.01<br>(0.00) |
| Intercept | 30.53***<br>(0.77) | 31.27***<br>(0.70) | 30.40***<br>(1.82) | 29.00***<br>(1.12) |

****p<0.001; **p<0.01; *p<0.05*